\documentclass[sigconf]{acmart}
\acmConference[EASE 2026]{The 30th International Conference on Evaluation and Assessment in Software Engineering}{10–13 June, 2026}{Glasgow, Scotland, United Kingdom}
\AtBeginDocument{%
  }

\usepackage{array}
\usepackage{multirow}
\usepackage{mdframed}
\usepackage{subcaption}
\usepackage{stfloats}
\usepackage{placeins}
\usepackage[table]{xcolor}
\usepackage{colortbl}
\usepackage{listings}
\usepackage{xcolor}
\usepackage{siunitx}
\usepackage[T1]{fontenc}

\makeatletter
\DeclareTextFontCommand{\texttt}{\ttfamily\hyphenchar\font=`\-}
\makeatother

\definecolor{codegreen}{rgb}{0,0.6,0}
\definecolor{codegray}{rgb}{0.5,0.5,0.5}
\definecolor{codepurple}{rgb}{0.58,0,0.82}
\definecolor{backcolour}{rgb}{0.95,0.95,0.92}

\NewEnviron{opus-box}[1][]{
 \vspace{0.1cm}
    \begin{mdframed}[backgroundcolor=gray!25]
        \BODY
    \end{mdframed}
    \vspace{0.1cm}
}

\NewEnviron{prompt-box}[1][]{
 \vspace{0.1cm}
    \begin{mdframed}[backgroundcolor=yellow!25]
        \BODY
    \end{mdframed}
    \vspace{0.1cm}
}

\definecolor{bestgreen}{RGB}{198,239,206}

\setcopyright{none}

\begin{document}

\title[Interprocedural Vulnerability Detection in Multiple Languages]{Interprocedural Vulnerability Detection in Multiple Languages: \\Assessing Effectiveness and Cost of Modern LLMs}

\title[Vulnerability Detection with Inteprocedural Context in Multiple Languages]{Vulnerability Detection with Interprocedural Context in Multiple Languages: Assessing Effectiveness and Cost of Modern LLMs}

\author{Kevin Lira}
\affiliation{%
  \institution{North Carolina State University}
  \city{Raleigh}
  \state{NC}
  \country{USA}
}
\email{kwlira@ncsu.edu}

\author{Baldoino Fonseca}
\affiliation{%
  \institution{Federal University of Alagoas}
  \city{Maceió}
  \state{AL}
  \country{Brazil}}
\email{baldoino@ic.ufal.br}

\author{Davy Baía}
\affiliation{%
  \institution{Federal University of Alagoas}
  \city{Maceió}
  \state{AL}
  \country{Brazil}}
\email{davy.baia@penedo.ufal.br}

\author{Márcio Ribeiro}
\affiliation{%
  \institution{Federal University of Alagoas}
  \city{Maceió}
  \state{AL}
  \country{Brazil}}
\email{marcio@ic.ufal.br}

\author{Wesley K. G. Assunção}
\affiliation{%
  \institution{North Carolina State University}
  \city{Raleigh}
  \state{NC}
  \country{USA}
}
\email{wguezas@ncsu.edu}

\renewcommand{\shortauthors}{Lira et al.}

\begin{abstract}
Large Language Models (LLMs) have been a promising way for automated vulnerability detection. However, most prior studies have explored the use of LLMs to detect vulnerabilities only within single functions, disregarding those related to interprocedural dependencies. These studies overlook vulnerabilities that arise from data and control flows that span multiple functions. Thus, leveraging the context provided by callers and callees may help identify vulnerabilities. This study empirically investigates the effectiveness of detection, the inference cost, and the quality of explanations of four modern LLMs (Claude Haiku 4.5, GPT-4.1 Mini, GPT-5 Mini, and Gemini 3 Flash) in detecting vulnerabilities related to interprocedural dependencies. To do that, we conducted an empirical study on 509 vulnerabilities from the ReposVul dataset, systematically varying the level of interprocedural context (target function code-only, target function + callers, and target function + callees) and evaluating the four modern LLMs across C, C++, and Python. The results show that Gemini 3 Flash offers the best cost-effectiveness trade-off for C vulnerabilities, achieving F1$~\geq~0.978$ at an estimated cost of \$0.50--\$0.58 per configuration, and Claude Haiku 4.5 correctly identified and explained the vulnerability in 93.6\% of the evaluated cases. Overall, the findings have direct implications for the design of AI-assisted security analysis tools that can generalize across codebases in multiple programming languages.

\end{abstract}


\maketitle

\section{Introduction}



Software vulnerabilities remain one of the most critical threats to modern software systems, enabling attacks ranging from data breaches to remote code execution \cite{li24}. Consequently, accurate and scalable vulnerability detection has become a central problem in software development \cite{li2021sysevr}. 
Large Language Models (LLMs) have been increasingly employed for automated software vulnerability detection~\cite{zhou25, khare2025understanding, germano25, saimbi24}. However, most previous studies focus on detecting vulnerabilities within single functions~\cite{fu2023chatgpt, ding2024primevul, zhou25}, disregarding vulnerabilities related to interprocedural dependencies~\cite{li24}.
These studies overlook security flaws that only manifest through data or control flow across multiple functions, not within a single function in isolation. 
This limitation is particularly relevant for vulnerability classes such as \emph{use-after-free}, buffer overflows triggered by externally controlled control flow, and injection vulnerabilities mediated by caller-supplied arguments, all of which require cross-function reasoning to be fully understood.

Another limitation of the literature is that studies predominantly target a single programming language for vulnerability detection, thereby neglecting the heterogeneity of programming languages employed in modern software development \cite{lira2025beyond}. Studying vulnerabilities across multiple languages is therefore essential to understand their prevalence, characteristics, and detectability in software development. 
Moreover, beyond detection effectiveness, the economic costs associated with LLM use and the explainability of their generated output are crucial practical factors \cite{khare2025understanding, bommasani2022}: black-box vulnerability detection without explainability limits usability, and methods that incur significant computational or manual analysis overhead are difficult to scale in real-world development pipelines. These challenges highlight the need for vulnerability-detection methods that leverage interprocedural information, are generalizable across multiple languages, and are cost-effective in practice.


In this paper, we conduct an empirical evaluation of four modern LLMs for interprocedural vulnerability detection. In particular, we evaluate the effectiveness and cost of applying the LLMs Claude Haiku~4.5, GPT-4.1 Mini, GPT-5 Mini, and Gemini~3 Flash for interprocedural vulnerability detection in existing software projects in C, C++, and Python. To do that, we extract 509 interprocedural vulnerabilities from the ReposVul dataset~\cite{wang2024repository}. We assess the effectiveness of the models in detecting these vulnerabilities, in terms of accuracy and F-measure, under three distinct input configurations: (i) we provide only the source code of the target function to be classified as vulnerable or not; (ii) we provide the target function along with information about the functions it invokes (i.e., callees); and (iii) we provide the target function together with information about the functions that invoke it (i.e., callers). This variation in input configurations enables us to evaluate whether providing more contextual information to the LLMs affects their effectiveness. We also assess the economic cost associated with the effectiveness of the LLMs. Finally, we instrument the LLMs to explain their classification decisions, to understand the rationale behind labeling each target function as either vulnerable or non-vulnerable. 

Interestingly, the results indicate that interprocedural context does not consistently improve LLM-based vulnerability detection and, in several cases, significantly degrades performance. However, this degradation is language-dependent. For C, GPT-4.1 Mini and GPT-5 Mini lose up to 25 and 11 percentage points of accuracy, respectively, when the caller or callee context is added, while Claude Haiku 4.5 and Gemini 3 Flash remain invariable across all context configurations and languages. From a cost perspective, adding interprocedural context nearly doubles token consumption and inference cost without delivering proportional accuracy gains. Gemini~3 Flash provides the best cost--performance trade-off for C vulnerabilities, achieving F1~$\geq$~0.978 at an estimated cost of \$0.50--\$0.58 per configuration. Claude Haiku 4.5 correctly identified and explained the vulnerability in 93.6\% of the evaluated cases. In contrast, the GPT models exhibit zero-score rates of correctness and comprehensiveness that are five times higher than Haiku 4.5 and Gemini 3 Flash.

\section{Motivating Example}
\label{sec:motivation_example}

\sloppy
Figures~\ref{fig:callee} and~\ref{fig:caller} depict a vulnerability (CVE-2016-10129)~\cite{CVE201610129} associated with a NULL pointer dereference in the \textit{Git Smart Protocol} implementation of \texttt{libgit2} (a portable C implementation of Git core functionality) in versions prior to 0.25.1. Identifying this vulnerability required analyzing two functions in distinct source files. Figure~\ref{fig:caller} presents the source code in \texttt{smart\_protocol.c}, which defines the function \texttt{add\_push\_report\_sideband\_pkt} (hereafter referred to as the \textit{caller} function). This function invokes \texttt{git\_pkt\_parse\_line} (hereafter referred to as the \textit{callee} function), implemented in \texttt{smart\_pkt.c}, as illustrated in Figure~\ref{fig:callee}.




\begin{figure}[!tp]
\centering
\begin{lstlisting}[language=C]
int git_pkt_parse_line(git_pkt **head, const char *line, const char **out, size_t bufflen)
{
    int32_t len = parse_len(line);

    ...

    if (len == PKT_LEN_SIZE) {
        *head = NULL;
        *out  = line;
        return 0;
    }

    ...
}
\end{lstlisting}
\vspace{-5mm}
\caption{Vulnerable callee (\texttt{smart\_pkt.c}). The empty-packet branch silently sets \texttt{*head = NULL} and returns success.}
\label{fig:callee}
\end{figure}


\begin{figure}[!tp]
\centering
\begin{lstlisting}[language=C]
static int add_push_report_sideband_pkt(git_push *push, git_pkt_data *data_pkt, git_buf *data_pkt_buf)
{
    git_pkt *pkt;
    const char *line = data_pkt->data;
    const char *line_end = line + data_pkt->len;
    int error;

    ...

    while (line < line_end) {
        error = git_pkt_parse_line(&pkt, line, &line, line_end - line);

        if (error < 0) goto done;

        /* pkt is not checked for NULL */
        if (pkt->type == GIT_PKT_ERR) {
            push->unpack_ok = 0;
            goto done;
        }

        if (pkt->type != GIT_PKT_DATA) {
            git_pkt_free(pkt);
            continue;
        }

        ...
    }
}
\end{lstlisting}
\vspace{-5mm}
\caption{Exploitable caller (\texttt{smart\_protocol.c}). A \texttt{NULL} pointer delivered on a success return is dereferenced without a null guard.}
\label{fig:caller}
\end{figure}

When examining the callee function \texttt{git\_pkt\_parse\_line} in isolation, we observe that it both returns \texttt{0} and assigns \texttt{*head = NULL}. However, without analyzing how the return value and the output parameter are interpreted by its caller (the function \texttt{add\_push\_report\_sideband\_pkt} in Figure~\ref{fig:caller}), it is not possible to determine whether this behavior is unsafe. The callee itself does not dereference a NULL pointer, and the assignment may plausibly serve as an intentional sentinel value designed for safe handling by the caller. The vulnerability arises solely from the interprocedural interaction between these two functions (caller and callee).


\section{Study Design}

In this study, we investigate the effectiveness of four modern LLMs, along with their associated economic costs, in detecting vulnerabilities in C, C++, and Python providing interprocedural contextual. To guide our investigation, we follow the research questions (RQs) described below:


\textbf{RQ1. \textit{How effective are LLMs in detecting vulnerabilities with interprocedural context?}} 
This RQ investigates the effectiveness of the LLMs Claude Haiku~4.5, GPT-4.1 Mini, GPT-5 Mini, and Gemini~3 Flash for vulnerability detection with interprocedural context in software projects in C, C++, and Python. We evaluate the LLM's effectiveness by considering three different variations of context: (i) providing only the target function code; (ii) providing the target function along with information about the callees; and (iii) providing the target function along with information about the callers. In this manner, we seek to systematically examine the impact of supplying additional interprocedural context on the effectiveness of the four LLMs.

\textbf{RQ2. \textit{To what extent do multiple languages influence model effectiveness?}} 
Programming languages differ in syntactic verbosity, structural patterns, and predominant vulnerability types. In our study, we focus on C, C++, and Python, which differ in abstraction level, memory management, and type systems, influencing how vulnerabilities manifest. C and C++ rely on manual memory management and limited runtime checks, making them prone to memory-safety issues such as buffer overflows and use-after-free errors~\cite{ritchie1978c}. C++ adds complexity through object lifetimes and resource management ~\cite{stroustrup2013c++}. In contrast, Python’s high-level design and automatic memory management mitigate low-level memory corruption but make it more susceptible to higher-level issues such as injection flaws and logic errors~\cite{lutz2001programming}.
Thus, this RQ investigates the influence of multiple languages on the LLM's effectiveness.

\textbf{RQ3. \textit{How efficient (in terms of cost) are LLMs in detecting interprocedural vulnerabilities in multiple languages?}}
Besides effectiveness (the focus of RQ2), differences across multiple languages may require LLMs to process more or fewer tokens per input and output, influencing LLM efficiency. By efficiency, we mean the relation between the effectiveness and economic cost for each LLM. Considering the cost of an LLM for vulnerability detection is important because modern models can incur substantial computational and financial overhead, affecting scalability and practical deployment in real-world development pipelines~\cite{wu2022sustainable}. A cost-aware evaluation ensures that improvements in detection effectiveness are balanced against resource consumption. This RQ investigates the efficiency of the models in multiple languages. 

\textbf{RQ4. \textit{To what extent do LLMs provide correct and complete explanations for the vulnerabilities they identify?}} 
Binary classification alone (e.g., vulnerable or non-vulnerable) is insufficient to support developer decision-making~\cite{guo2018lemna, li2021vulnerability}: a security alert must also identify the vulnerability type, its location in the code, and the potential exploitation vector. This RQ involves a qualitative analysis of the explanations generated by each model, assessing whether correct classifications are accompanied by technically accurate and complete explanations.

\subsection{Dataset}

We used the ReposVul~\cite{wang2024repository} dataset as the source of information for this study. ReposVul is a public repository-level vulnerability dataset constructed from patch commits collected from the NVD (\textit{National Vulnerability Database}) CVE records. The dataset contains 6,134 CVE entries, distributed across 1,491 open-source projects and covering 236 CWE types in C, C++, and Python. To ensure annotation quality, ReposVul employs an LLM-based \textit{vulnerability untangling} module combined with static analysis tools to distinguish vulnerability-repair changes from unrelated modifications in mixed (i.e., \textit{tangled}) patches. Each dataset entry includes CVE metadata, patch commit information, the function code before and after the patch, and annotations at multiple levels of granularity (line, function, file, and repository).

From the original dataset, we applied a two-stage filtering process to construct the subset used in our evaluation. First, we kept only entries that contained at least one vulnerable function. Second, for each extracted function, we identified interprocedural caller-callee relationships using Abstract Syntax Tree (AST) construction.

\subsection{LLM Instrumentation}

We selected four LLMs from three major providers: Anthropic, OpenAI, and Google.  Table~\ref{tab:llm_list} lists the models used in this study. The selection criteria prioritized models that support long-context input to accommodate interprocedural context blocks and that represent distinct cost-performance profiles within each provider's offering, enabling the cost-effectiveness analysis conducted in RQ3.

\begin{table}[!tp]
  \centering
  \caption{LLMs evaluated in this study and their pricing per 1M tokens (March 2026).}
  \label{tab:llm_list}
  \addtolength{\tabcolsep}{-2pt}
  \begin{tabular}{llrr}
    \toprule
    \textbf{Model} & \textbf{Provider} &
    \multicolumn{1}{c}{\textbf{Input \small{(USD)}}} &
    \multicolumn{1}{c}{\textbf{Output \small{(USD)}}} \\
    \midrule
    Claude Haiku 4.5~\cite{haiku45} & Anthropic & 1.00 & 5.00 \\
    Gemini 3 Flash~\cite{gemini3} & Google & 0.50 & 3.00 \\
    GPT 4.1 Mini~\cite{gpt41} & OpenAI & 0.40 & 1.60 \\
    GPT 5 Mini~\cite{gpt51} & OpenAI & 0.25 & 2.00 \\
    \bottomrule
  \end{tabular}
\end{table}

\subsubsection{Prompt engineering}

For each function in the dataset, we generated prompt variations corresponding to each available context level. All prompts follow a fixed structure composed of: (i) a task instruction requesting the binary classification into vulnerable (1) or non-vulnerable (0) and a structured explanation; and (ii) the interprocedural context block. The interprocedural context block constitutes the primary independent variable of the study, assuming three possible configurations:

\begin{itemize}
    \item \textbf{Code Only (CO)}: the code of the vulnerable function, without any information about related functions.
    \item \textbf{Code + Callees (CC)}: the code of the function augmented with the functions it directly invokes.
    \item \textbf{Code + Callers (CK)}: the code of the function augmented with the functions that directly invoke it.
\end{itemize}

We queried all models via their official APIs, setting the temperature to 0 and the number of output tokens to 4096 to minimize response variability and ensure reproducibility. We generate each prompt variation by combining each function with a given additional context level. The prompt template used across all configurations is shown in Figure~\ref{fig:prompt_template}. The placeholders \texttt{<function code>} and \texttt{<callees|callers code>} are substituted according to the active configuration. The prompt enforces a fixed output format, enabling automated extraction of the classification and the explanation.

\begin{figure}[b]
\begin{prompt-box}
{\fontsize{7}{8}\selectfont
\begin{minipage}{\linewidth}
You are a security code reviewer performing vulnerability detection. Analyze the function below and determine whether it contains a security vulnerability. Please provide your response in the following format: \\

1. EXPLANATION: Explain where the vulnerability is located in the code. Be specific about:

- Which lines or code sections are vulnerable \\
- What type of vulnerability it is (e.g., buffer overflow, SQL injection, use-after-free, etc.) \\
- Why this code is vulnerable \\
- If no vulnerability exists, explain why the code is secure \\

2. CLASSIFICATION: After your explanation, provide a binary classification on a new line:

- 1 if a vulnerability is present \\
- 0 if no vulnerability is present \\

Format your response exactly as:

EXPLANATION: [your detailed explanation here] \\
CLASSIFICATION: [0 or 1] \\

<function code>

<callees|callers code>
\end{minipage}
}
\end{prompt-box}
    \caption{Prompt template used.}
    \label{fig:prompt_template}
\end{figure}

We automatically processed model responses using a regular-expression-based parser, flagged outputs that did not conform to the expected format, and excluded them from the quantitative analysis. We also preserved the raw content of the model response for the qualitative assessment performed in RQ4. We derived the ground-truth labels from the \texttt{target} field in the ReposVul dataset, where \texttt{target~=~1} denotes a vulnerable function and \texttt{target~=~0} indicates non-vulnerable code.

\subsection{Metrics}
To evaluate the LLMs' vulnerability-detection capability, we used standard binary classification metrics, as follows: 

\textbf{Accuracy (A)} represents the proportion of correctly classified snippets relative to the total number of evaluated snippets, defined as $A = \frac{TP + TN}{TP + TN + FP + FN}$. For classification, we mapped the LLM outputs to two possible labels: vulnerable (1) and non-vulnerable (0). Since the evaluation subset contains only vulnerable functions (\texttt{target~=~1}), there are no true negatives or false positives in this study. The equation therefore reduces to $A = \frac{TP}{TP + FN}$, making accuracy equivalent to recall throughout all reported results. Precision equals $1.0$ under all conditions by construction.

\textbf{Precision (P)}, defined as $P = \frac{TP}{TP + FP}$, is the ratio between true positive predictions and the total number of instances classified as positive, measuring the reliability of the LLM vulnerability alerts. 

\textbf{Recall (R)}, $R = \frac{TP}{TP + FN}$, is the ratio between true positive predictions and the total number of positive instances, capturing the LLM's ability to identify existing vulnerabilities without omission. 

The \textbf{F1-score (F1)}, computed as the harmonic mean of precision and recall, is given by $F1 = 2 \cdot \frac{P \cdot R}{P + R}$.

\subsection{Data Analysis}
\label{sec:data_analysis}

To determine whether accuracy differences across context configurations are statistically meaningful, we apply \textit{McNemar's test}~\cite{mcnemar1947note}, a non-parametric paired test designed to compare two classifiers on the same set of observations. For each LLM and each pair of context configurations (CO\,vs.\,CC, CO\,vs.\,CK, and CC\,vs.\,CK), we construct a $2 \times 2$ contingency table over the subset of functions shared by both configurations, i.e., functions for which both prompt types can be generated. This paired design controls for sample-level variation and isolates the effect of the context manipulation.

Each model produces three pairwise comparisons; thus, we apply the \textit{Holm--Bonferroni correction} to control the family-wise error rate within each model. We rank comparisons by ascending $p$-value and derive an adjusted significance threshold $\alpha$ for each test. A comparison is statistically significant when its raw $p$-value falls below its corresponding corrected threshold.

To quantify the magnitude of observed differences, we compute the \textit{phi coefficient} ($\Phi$), the standard effect-size measure for $2 \times 2$ contingency tables. By convention, $|\Phi| < 0.1$ indicates a negligible effect, $0.1 \leq |\Phi| < 0.3$ a small effect, $0.3 \leq |\Phi| < 0.5$ a medium effect, and $|\Phi| \geq 0.5$ a large effect~\cite{cohen1988statistical}. A positive $\Phi$ indicates that the first configuration in the pair outperforms the second, whereas a negative value indicates the opposite.


We estimate inference cost from the prompt token counts collected during evaluation. For each function and context configuration, we count input tokens before prompting each model and compute the estimated cost using the cost-per-million (CPM) prices published by each provider at the time of the study. This procedure produces per-language, per-configuration cost estimates used in RQ3 to characterize each model's cost--performance trade-off.

To answer RQ4, we manually evaluate the explanations generated by the models for the detected vulnerabilities. From the set of analyzed functions, we sampled 251 explanations using a stratified sampling scheme by programming language, ensuring a 95\% confidence level with a 5\% margin of error within each stratum: 173 samples for C $(N = 310)$, 63 for Python $(N = 75)$, and 15 for C++ (N = 16). For each sampled function, we evaluated the explanations produced by each model under the Code-Only (CO) configuration, yielding a total of 1,004 manual assessments.

We evaluated the explanations along two independent dimensions: (i) \textit{correctness}, which assesses whether the explanation correctly identifies the vulnerability type and its location in the code; and (ii) \textit{comprehensiveness}, which evaluates whether the LLM's explanation of its decision provides sufficient information to guide a developer in fixing the issue, including the exploitation vector or a repair suggestion. Both dimensions were scored on a 0–2 scale, as defined in Tables~\ref{tab:rubric-accuracy} and~\ref{tab:rubric-comprehensiveness}. The ground truth used as a reference for each evaluation consisted of CVE metadata from ReposVul, including vulnerability type, affected lines, and patch description.

\begin{table}[t]
\caption{Rubric for Correctness Evaluation}
\label{tab:rubric-accuracy}
\centering
\begin{tabular}{c p{0.8\columnwidth}}
\toprule
\textbf{Score} & \textbf{Criteria} \\
\midrule
2 & Correct vulnerability type, correct location in code, and correct root cause. \\
\midrule
1 & Correct type but wrong location, or incomplete root cause. \\
\midrule
0 & Wrong type or fundamentally incorrect explanation. \\
\bottomrule
\end{tabular}
\end{table}

\begin{table}[t]
\caption{Rubric for Comprehensiveness Evaluation}
\label{tab:rubric-comprehensiveness}
\centering
\begin{tabular}{c p{0.8\columnwidth}}
\toprule
\textbf{Score} & \textbf{Criteria} \\
\midrule
2 & Explains type, location, and root cause, and includes the exploitation vector/repair suggestion. \\
\midrule
1 & Explains type and location but omits the exploitation vector/repair suggestion. \\
\midrule
0 & Generic explanation with no code-specific detail. \\
\bottomrule
\end{tabular}
\end{table}

\subsection{Replication Package}

The replication package~\cite{supplementary} for this study, which includes all resources necessary to reproduce the results, is publicly available on GitHub. This package contains the code used, the LLM output generated during the experiments, and the filtered vulnerability dataset used for testing. The replication package also includes a visualization dashboard and detailed instructions for setting up the environment and executing the scripts, enabling accurate reproduction of the process used in this research.

\section{Results}

This section presents the results and analysis of our study.

\begin{table*}[!tp]
\centering
\caption{McNemar test results (across all programming languages).}
\label{tab:mcnemar_overall}
\begin{tabular}{lcccccccccc}
\toprule
\textbf{Model} & \textbf{Context Comparison} & \textbf{N} & $\boldsymbol{\chi^2}$ & 
\textbf{p} & $\boldsymbol{\alpha}$ & \textbf{Sig.} & 
\textbf{Acc$_1$} & \textbf{Acc$_2$} & $\boldsymbol{\Phi}$ & \textbf{Effect} \\
\midrule
\multirow{3}{*}{Claude Haiku 4.5}
  & CO -- CC & 237 & 0.571 & 0.4497 & 0.0100 & -- & 0.9916 & 0.9789 & 0.074 & Small \\
  & CO -- CK & 221 & 0.083 & 0.7728 & 0.0250 & -- & 0.9683 & 0.9774 & -0.039 & Small \\
  & CC -- CK & 129 & 0.000 & 1.0000 & 0.0500 & -- & 0.9690 & 0.9767 & -0.039 & Small \\
\midrule
\multirow{3}{*}{Gemini 3 Flash}
  & CO -- CC & 157 & 5.143 & 0.0233 & 0.0056 & -- & 1.0000 & 0.9554 & 0.211 & Medium \\
  & CO -- CK & 140 & 1.500 & 0.2207 & 0.0083 & -- & 0.9929 & 0.9643 & 0.138 & Medium \\
  & CC -- CK & 68 & 0.500 & 0.4795 & 0.0125 & -- & 0.9559 & 0.9265 & 0.171 & Medium \\
\midrule
\multirow{3}{*}{GPT-4.1 Mini}
  & CO -- CC & 237 & 37.123 & 0.0000 & 0.0042 & $\checkmark$ & 0.7384 & 0.5401 & 0.404 & Large \\
  & CO -- CK & 221 & 0.103 & 0.7488 & 0.0167 & -- & 0.7059 & 0.6923 & 0.032 & Small \\
  & CC -- CK & 129 & 22.781 & 0.0000 & 0.0045 & $\checkmark$ & 0.4651 & 0.6822 & -0.436 & Large \\
\midrule
\multirow{3}{*}{GPT-5 Mini}
  & CO -- CC & 222 & 4.654 & 0.0310 & 0.0063 & -- & 0.8964 & 0.8423 & 0.158 & Medium \\
  & CO -- CK & 213 & 13.081 & 0.0003 & 0.0050 & $\checkmark$ & 0.8779 & 0.7700 & 0.259 & Medium \\
  & CC -- CK & 122 & 1.895 & 0.1687 & 0.0071 & -- & 0.8361 & 0.7787 & 0.145 & Medium \\
\bottomrule
\end{tabular}

{\footnotesize $\chi^2$ = test statistic; $p$ = raw p-value; $\alpha$ = Holm-Bonferroni corrected significance; \textit{Sig.} = significance after correction; $\text{Acc}_1$ and $\text{Acc}_2$ = the accuracy of the first and\\second context in the pair; $\Phi$ = effect size, where $\Phi > 0$ indicates that the first context outperforms the second, and $\Phi < 0$ the opposite.}
\end{table*}

\subsection{RQ1. How effective are LLMs in detecting vulnerabilities with interprocedural context?}

This RQ examines whether increasing the amount of interprocedural context yields a measurable impact over the CO baseline. 

Table~\ref{tab:overall_rq1} reports the accuracy, true positives (TP), and F1-score for each model under each context configuration for all programming languages. The sample size ($N$) varies by configuration, as the extraction of callers and callees depends on the interprocedural information in the call graph, which does not exist for all functions. 

\begin{table}[!tp]
\centering
\caption{Overall accuracy and F1-score per model and context configuration across all programming languages.}
\label{tab:overall_rq1}
\resizebox{\columnwidth}{!}{%
\begin{tabular}{lccccc}
\toprule
\textbf{Model} & \textbf{Context} & \textbf{N} & 
\textbf{TP} & \textbf{Accuracy} & \textbf{F1} \\
\midrule
\multirow{3}{*}{Claude Haiku 4.5}
  & CO & 451 & 440 & \cellcolor{bestgreen}\textbf{0.9756} & \cellcolor{bestgreen}\textbf{0.9877} \\
  & CC & 237 & 232 & 0.9789 & 0.9893 \\
  & CK & 221 & 216 & 0.9774 & 0.9886 \\
\midrule
\multirow{3}{*}{Gemini 3 Flash}
  & CO & 408 & 402 & \cellcolor{bestgreen}\textbf{0.9853} & \cellcolor{bestgreen}\textbf{0.9926} \\
  & CC & 170 & 162 & 0.9529 & 0.9759 \\
  & CK & 149 & 143 & 0.9597 & 0.9795 \\
\midrule
\multirow{3}{*}{GPT-5 Mini}
  & CO & 442 & 401 & \cellcolor{bestgreen}\textbf{0.9072} & \cellcolor{bestgreen}\textbf{0.9514} \\
  & CC & 230 & 192 & 0.8348 & 0.9100 \\
  & CK & 217 & 168 & 0.7742 & 0.8727 \\
\midrule
\multirow{3}{*}{GPT-4.1 Mini}
  & CO & 451 & 341 & \cellcolor{bestgreen}\textbf{0.7561} & \cellcolor{bestgreen}\textbf{0.8611} \\
  & CC & 237 & 128 & 0.5401 & 0.7014 \\
  & CK & 221 & 153 & 0.6923 & 0.8182 \\
\bottomrule
\end{tabular}%
}
\end{table}

The results in Table~\ref{tab:overall_rq1} show a clear and consistent pattern across all models: the CO configuration achieves the best overall performance, both in terms of accuracy and F1-score. For Claude Haiku~4.5 and Gemini 3 Flash, CO yields the highest Accuracy (0.9756 and 0.9853, respectively) and near-perfect F1-scores (0.9877 and 0.9926). Although CC and CK remain competitive for these two models, they do not surpass CO. The difference becomes more pronounced for GPT-5 Mini and especially GPT-4.1 Mini. GPT-5 Mini drops from 0.9072 accuracy in CO to 0.7742 in CK, while GPT-4.1 Mini shows a dramatic decline from 0.7561 in CO to 0.5401 in CC. 
When comparing models, a clear performance hierarchy emerges. Gemini 3 Flash achieves the highest overall effectiveness, reaching 0.9853 accuracy and 0.9926 F1-score under CO, and maintaining strong results across other contexts. Claude Haiku 4.5 follows closely, with similarly high and stable performance across CO, CC, and CK. GPT-5 Mini performs moderately well under CO but exhibits noticeable sensitivity to context changes, particularly under CK. GPT-4.1 Mini performs the weakest overall and is the most affected by context variation, with substantial degradation under CC and CK. 


Table~\ref{tab:mcnemar_overall} presents the results of McNemar's test per model for all context comparisons across all programming languages. Each row corresponds to a pair of context configurations evaluated on the subset of functions that appear in both configurations (i.e., paired samples, reported as $N$). 
Language-specific McNemar tables for all configurations are provided in our supplementary material~\cite{supplementary}.

The statistical analysis confirms that contextual configuration has model-dependent effects. Stronger LLMs (Claude Haiku 4.5 and Gemini 3 Flash) show limited or moderate sensitivity to context, with mostly small-to-medium effects and few statistically significant differences. In contrast, smaller models (particularly GPT-4.1 Mini) exhibit large and statistically significant performance shifts across configurations, especially in CC. The results in Table~\ref{tab:mcnemar_overall} confirm that CO is the most stable and often superior configuration. 

\begin{opus-box}
\textbf{Answering RQ1:} Interprocedural context does not consistently improve LLM  accuracy in vulnerability detection. The CO setting consistently provides the best results, and larger LLMs demonstrate greater robustness and generalization across different contextual inputs. Of the 12 pairwise comparisons, only three reached statistical significance, and in every case the effect favors the configuration with \emph{less} context.

\end{opus-box}

\subsection{RQ2. To what extent do multiple languages influence model effectiveness?}

This RQ investigates whether LLM accuracy remains stable across languages.
Table~\ref{tab:overall_c} reports accuracy, true positives (TP), and F1-score for each LLM under the three context configurations (CO, CC, CK), considering only functions written in C. 

The C++ subset is too small for reliable cross-model comparison: after filtering for interprocedural availability, the three context configurations contain only 18, 6, and 5 functions for the largest model. We therefore exclude C++ from the statistical analysis and limit observations to descriptive statistics: all models achieve perfect or near-perfect accuracy under CO (Acc $\geq$ 0.889), but sample sizes collapse to as few as $N=1$ under CK, rendering any trend uninterpretable. C++ results are reported in our replication package for completeness.

\begin{table}[!tp]
\centering
\caption{Overall accuracy and F1-score per model and context configuration for C.}
\label{tab:overall_c}
\resizebox{\columnwidth}{!}{%
\begin{tabular}{lccccc}
\toprule
\textbf{Model} & \textbf{Context} & \textbf{N} &
\textbf{TP} & \textbf{Accuracy} & \textbf{F1} \\
\midrule
\multirow{3}{*}{Claude Haiku 4.5}
  & CO & 356 & 349 & \cellcolor{bestgreen}\textbf{0.9803} & \cellcolor{bestgreen}\textbf{0.9901} \\
  & CC & 204 & 200 & 0.9804 & 0.9901 \\
  & CK & 189 & 185 & 0.9788 & 0.9893 \\
\midrule
\multirow{3}{*}{Gemini 3 Flash}
  & CO & 316 & 314 & \cellcolor{bestgreen}\textbf{0.9937} & \cellcolor{bestgreen}\textbf{0.9968} \\
  & CC & 142 & 136 & 0.9577 & 0.9784 \\
  & CK & 125 & 121 & 0.9680 & 0.9837 \\
\midrule
\multirow{3}{*}{GPT-5 Mini}
  & CO & 348 & 315 & \cellcolor{bestgreen}\textbf{0.9052} & \cellcolor{bestgreen}\textbf{0.9502} \\
  & CC & 197 & 166 & 0.8426 & 0.9146 \\
  & CK & 185 & 147 & 0.7946 & 0.8855 \\
\midrule
\multirow{3}{*}{GPT-4.1 Mini}
  & CO & 356 & 269 & \cellcolor{bestgreen}\textbf{0.7556} & \cellcolor{bestgreen}\textbf{0.8608} \\
  & CC & 204 & 104 & 0.5098 & 0.6753 \\
  & CK & 189 & 130 & 0.6878 & 0.8150 \\
\bottomrule
\end{tabular}%
}
\end{table}


Table~\ref{tab:overall_python} presents accuracy, true positives (TP), and F1-score for Python functions. Unlike the previous tables, in some model–language combinations, the CC and CK configurations outperform CO; Numbers in boldface highlights the configuration that achieved the highest observed accuracy.

\begin{table}[!tp]
\centering
\caption{Overall accuracy and F1-score per model and context configuration for Python.}
\label{tab:overall_python}
\resizebox{\columnwidth}{!}{%
\begin{tabular}{lccccc}
\toprule
\textbf{Model} & \textbf{Context} & \textbf{N} &
\textbf{TP} & \textbf{Accuracy} & \textbf{F1} \\
\midrule
\multirow{3}{*}{Claude Haiku 4.5}
  & CO & 77 & 73 & 0.9481 & 0.9733 \\
  & CC & 27 & 26 & \cellcolor{bestgreen}\textbf{0.9630} & \cellcolor{bestgreen}\textbf{0.9811} \\
  & CK & 27 & 26 & \cellcolor{bestgreen}\textbf{0.9630} & \cellcolor{bestgreen}\textbf{0.9811} \\
\midrule
\multirow{3}{*}{Gemini 3 Flash}
  & CO & 75 & 71 & \cellcolor{bestgreen}\textbf{0.9467} & \cellcolor{bestgreen}\textbf{0.9726} \\
  & CC & 24 & 22 & 0.9167 & 0.9565 \\
  & CK & 22 & 21 & 0.9545 & 0.9767 \\
\midrule
\multirow{3}{*}{GPT-5 Mini}
  & CO & 77 & 69 & \cellcolor{bestgreen}\textbf{0.8961} & \cellcolor{bestgreen}\textbf{0.9452} \\
  & CC & 27 & 21 & 0.7778 & 0.8750 \\
  & CK & 27 & 18 & 0.6667 & 0.8000 \\
\midrule
\multirow{3}{*}{GPT-4.1 Mini}
  & CO & 77 & 56 & 0.7273 & 0.8421 \\
  & CC & 27 & 20 & \cellcolor{bestgreen}\textbf{0.7407} & \cellcolor{bestgreen}\textbf{0.8511} \\
  & CK & 27 & 19 & 0.7037 & 0.8261 \\
\bottomrule
\end{tabular}%
}
\end{table}

\vspace{2mm}
\begin{opus-box}
\textbf{Answering RQ2:} LLM performance varies across programming languages. In C, incorporating interprocedural context reduces the accuracy of the GPT models, with statistically significant differences observed for GPT-4.1 Mini and GPT-5 Mini models. In Python, a similar trend emerges, although without statistical significance. The results obtained for C++ are not statistically reliable due to the limited sample size.
\end{opus-box}

\subsection{RQ3. How efficient are LLMs in detecting interprocedural vulnerabilities in multiple languages?}

This RQ examines the efficiency (i.e., the relation between detection effectiveness and inference cost) of each model across programming languages. Although language-specific accuracy was previously reported in RQ2 (Tables~\ref{tab:overall_c} and~\ref{tab:overall_python}), this analysis integrates token consumption and estimated execution cost to characterize the cost-effectiveness of each model within each language.

\begin{figure*}
\centering
\begin{subfigure}{0.33\textwidth}
    \centering
    \includegraphics[width=\linewidth]{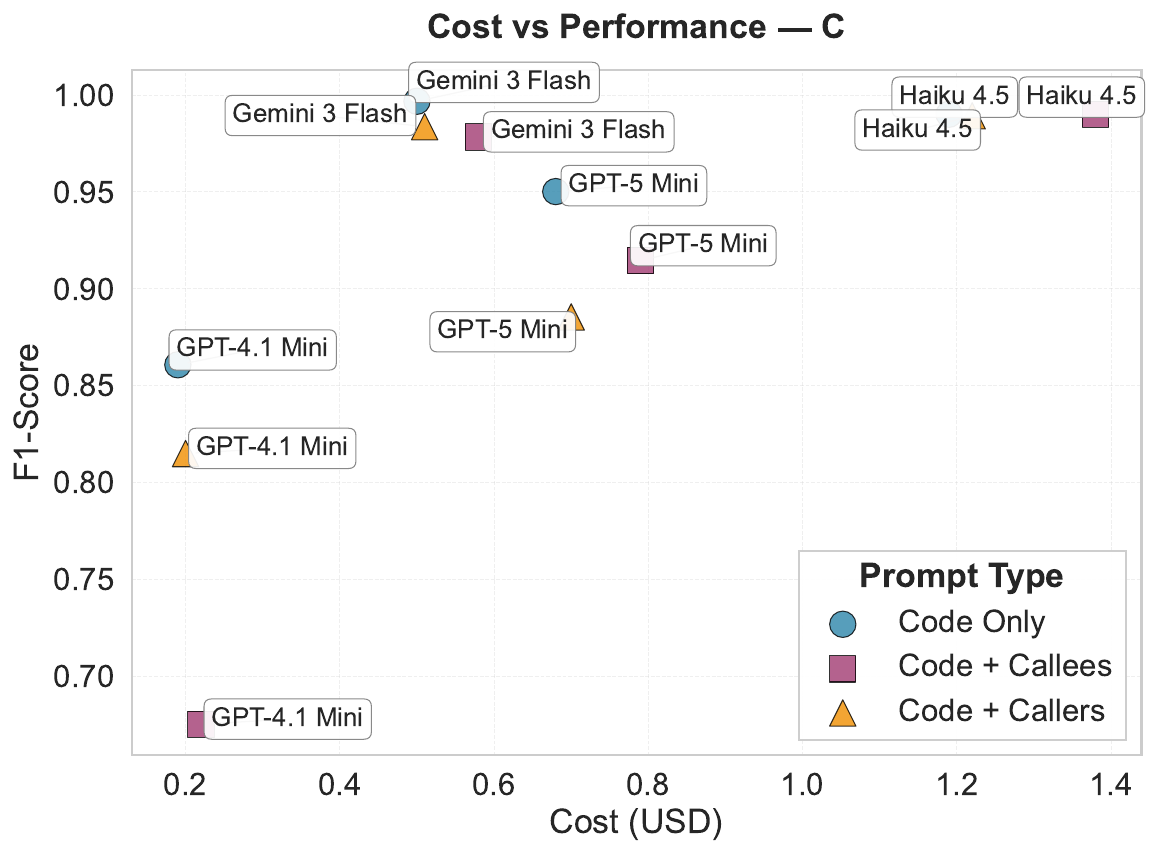}
    \caption{C}
\end{subfigure}
\hfill
\begin{subfigure}{0.33\textwidth}
    \centering
    \includegraphics[width=\linewidth]{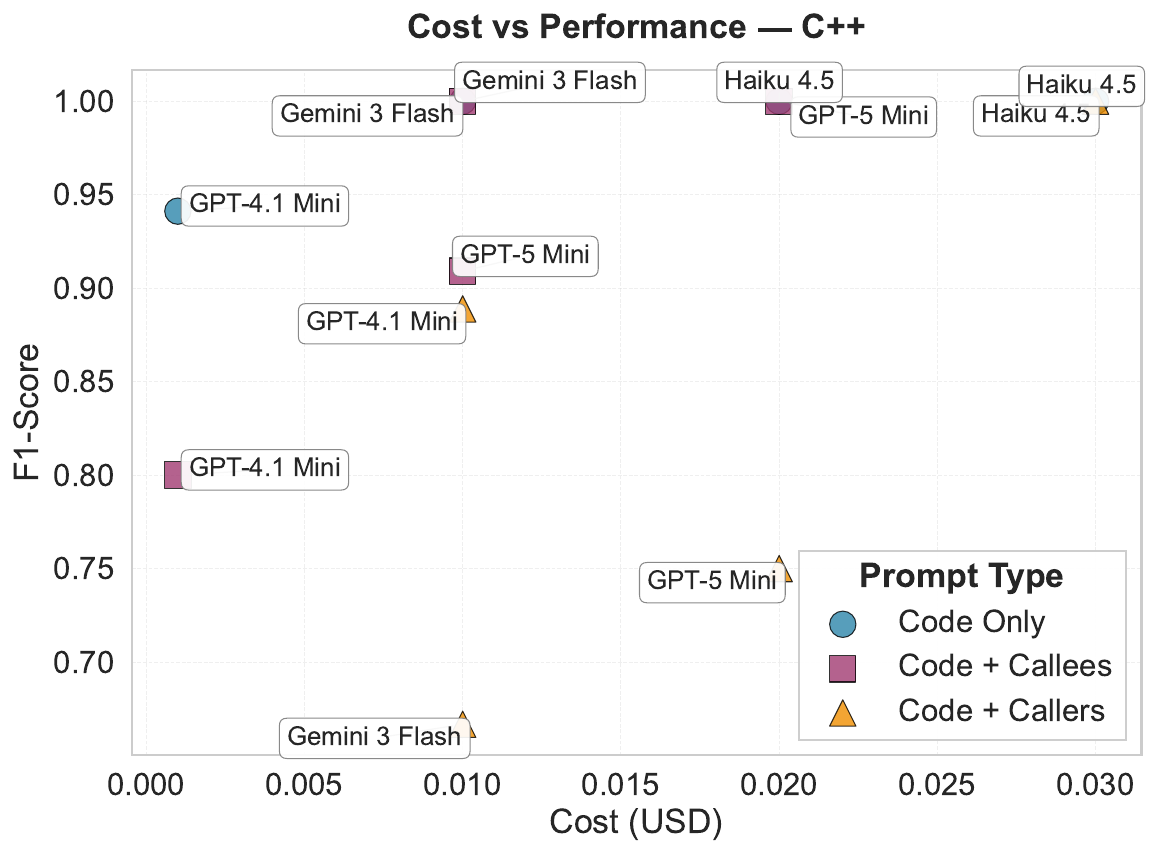}
    \caption{C++}
\end{subfigure}
\hfill
\begin{subfigure}{0.33\textwidth}
    \centering
    \includegraphics[width=\linewidth]{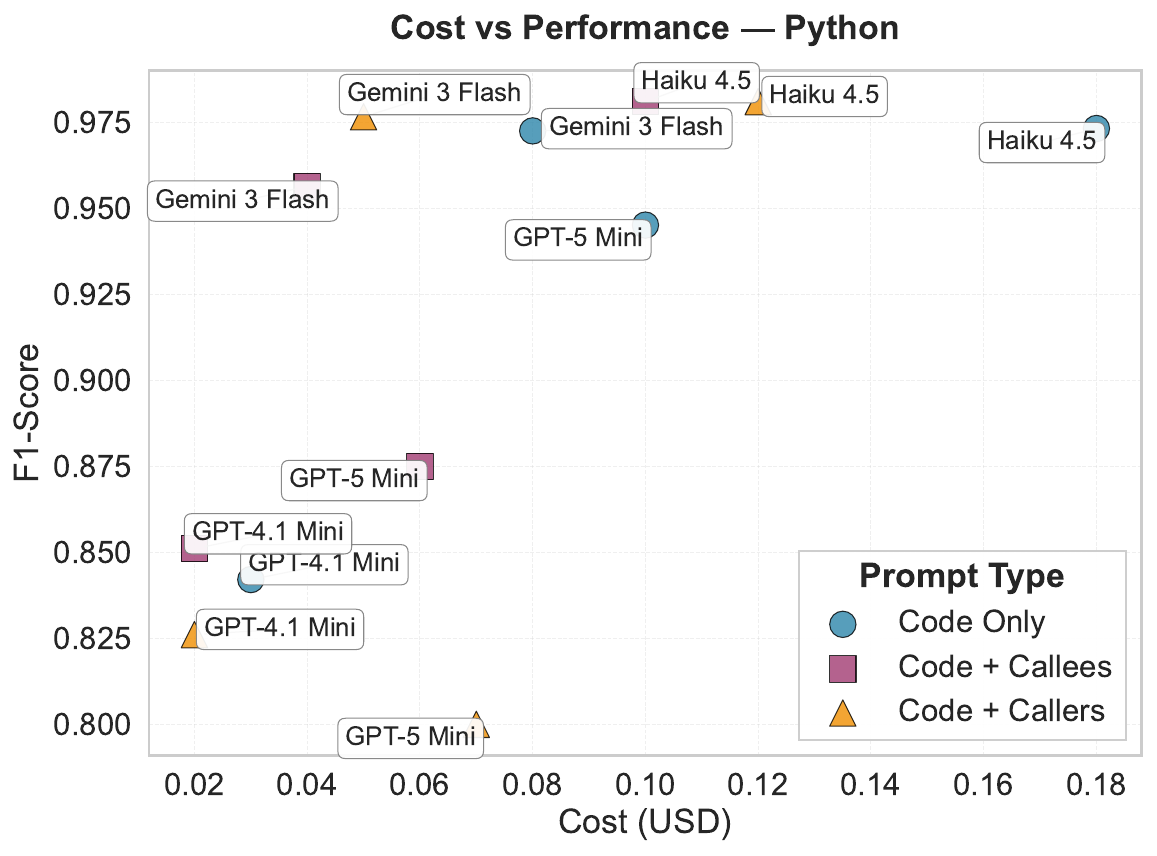}
    \caption{Python}
\end{subfigure}
\caption{Cost--performance trade-off across programming languages.}
\label{fig:cost_f1_all}
\end{figure*}

Figure~\ref{fig:cost_f1_all} plots the \textit{cost vs. performance} trade-off for each model and context configuration across the three programming languages. Cost is measured in USD based on official per-token pricing applied to the prompt token counts collected during evaluation; performance is measured by F1-score.

Token consumption varies across programming languages and prompt configurations. C prompts have a mean of 2,367 tokens under CO and approximately 4,785 and 4,586 tokens with CC and CK contexts, respectively. Adding interprocedural context nearly doubles prompt length in C. Python prompts are more compact (mean CO: 1,678 tokens). In comparison, C++ prompts are both fewer in number and shorter on average (mean CO: 1,183 tokens), resulting in lower absolute inference costs. Across the full evaluation, total spending ranged from \$0.87 for GPT-4.1 Mini to \$7.23 for Claude Haiku, with GPT-5 Mini and Gemini 3 Flash incurring intermediate costs of \$3.04 and \$3.76, respectively.

For C (Figure~\ref{fig:cost_f1_all}a), Gemini 3 Flash provides the most favorable cost–performance trade-off. It achieves near-perfect F1 scores ($\geq$0.978) at moderate per-configuration costs (\$0.50–\$0.58), placing it close to the efficiency frontier. Claude Haiku achieves slightly higher F1 scores ($\geq$0.985) but at more than double the cost (\$1.19--\$1.38), offering only a small improvement for the price. GPT-5 Mini occupies an intermediate position, achieving F1 = 0.950 under CO at \$0.68; performance declines under contextual configurations while cost increases. GPT-4.1 Mini remains the least expensive option (\$0.19--\$0.22). However, its performance degrades when callees (CC) are added (F1 decreases from 0.861 under CO to 0.675 under CC), reducing its overall cost-effectiveness in this language. The C++ results (Figure~\ref{fig:cost_f1_all}b) show absolute costs approximately two orders of magnitude lower than those observed for C, ranging from below \$0.01 to approximately \$0.03 per configuration. Claude Haiku and Gemini 3 Flash achieve perfect or near-perfect F1 under CO; however, the limited sample size constrains the reliability of cross-model cost-efficiency comparisons for this language.

For Python (Figure~\ref{fig:cost_f1_all}c), Claude Haiku achieves the highest F1 values ($\geq 0.973$ across configurations) at costs between \$0.10 and \$0.18. Gemini 3 Flash delivers comparable performance under CC and CK at lower cost (\$0.04--\$0.08), yielding a more favorable cost--performance balance. GPT-4.1 Mini remains the least expensive model (\$0.02--\$0.03) but also the least accurate (F1: 0.826--0.851). GPT-5 Mini shows the greatest sensitivity to added context, with F1 declining from 0.945 under CO to 0.800 under CK, suggesting that additional interprocedural information does not consistently improve detection quality.

\vspace{2mm}
\begin{opus-box} 
\textbf{Answering RQ3:} Inference cost scales with token volume, approximately doubling when interprocedural context is added. Across all languages, Gemini 3 Flash offers the best cost-performance balance, achieving near-perfect F1 in C. In no language does adding interprocedural context improve cost-effectiveness: higher token consumption is paired with equal or lower accuracy across models. 
\end{opus-box}

\subsection{RQ4. To what extent do LLMs provide correct and complete explanations for the vulnerabilities they identify?}

We targeted 1,004 assessments (251 functions × 4 models), of which 986 were evaluable; 18 responses were excluded due to the absence of a structured explanation. The final sample comprised 173 C functions, 63 Python functions, and 15 C++ functions, each evaluated across two dimensions: \textit{correctness} (correctness of vulnerability identification) and \textit{comprehensiveness} (quality and actionability of the LLM's explanation), both scored on a 0--2 scale (see Section~\ref{sec:data_analysis}).

Table~\ref{tab:overall-score-distribution} presents the distribution of correctness and comprehensiveness scores for each LLM across the manual evaluations. All models demonstrate strong performance on both dimensions, with overall means exceeding 1.83 on a 0–2 scale. Claude Haiku 4.5 achieved the best aggregate results, with mean scores of 1.956 for correctness and 1.928 for comprehensiveness, obtaining a perfect (2,2) score in 93.6\% of the evaluations.
Gemini 3 Flash exhibited the most asymmetric pattern across the two dimensions. While its correctness (1.954) is comparable to Claude Haiku 4.5, its comprehensiveness (1.814) is the lowest among the evaluated models, with 40 evaluations (16.9\%) receiving a partial score in this dimension. The GPT models showed the highest rates of zero-scores: 6.1\% for GPT-5 Mini and 6.8\% for GPT-4.1 Mini, compared to less than 1.2\% for Haiku and Gemini. This indicates that, although fewer, cases of incorrect vulnerability identification are more concentrated in the OpenAI models.

\begin{table}[t]
\caption{Overall score distribution (Correctness and Comprehensiveness) statistics.}
\label{tab:overall-score-distribution}
\centering
\begin{tabular}{l ccc ccc c}
\toprule
\textbf{Model} & \multicolumn{3}{c}{\textbf{Corr.}} & \multicolumn{3}{c}{\textbf{Comp.}} & \textbf{Perfect} \\
\cmidrule(lr){2-4} \cmidrule(lr){5-7}
 & 0 & 1 & 2 & 0 & 1 & 2 & (2,2) \\
\midrule
Claude Haiku 4.5       & 3  & 5 & 243 & 2  & 14 & 235 & \cellcolor{bestgreen}\textbf{235} \\
Gemini 3 Flash  & 2  & 7 & 228 & 2  & 40 & 195 & 194 \\
GPT-4.1 Mini    & 17 & 5 & 229 & 16 & 10 & 225 & 223 \\
GPT-5 Mini      & 15 & 4 & 228 & 14 & 10 & 223 & 223 \\
\midrule
\textbf{Total (N=986)} 
                 & 37 & 21 & 928 
                 & 34 & 74 & 878 
                 & 875 \\
\bottomrule
\end{tabular}
\end{table}

Table~\ref{tab:model-performance-language} presents the distribution of manual evaluations by programming language. For C, Claude Haiku 4.5 leads with an overall mean of 1.922, while Gemini 3 Flash exhibits the largest discrepancy between correctness and comprehensiveness among the four evaluated models.
In Python, Claude Haiku 4.5 achieves a mean score of 1.984 in both dimensions and zero score-0 evaluations. GPT-4.1 Mini shows the weakest performance in Python, with six evaluations receiving a score of 0 in both correctness and comprehensiveness. In C++, all models achieve perfect correctness (2.000), and three of the four also reach perfect comprehensiveness; only Gemini 3 Flash records a single partial score in comprehensiveness.

Overall, comprehensiveness is the more challenging measure across all models. While the proportion of maximum scores in correctness ranges from 91.2\% to 96.8\%, the corresponding rate for comprehensiveness varies from 82.3\% to 93.6\%. This pattern suggests that although LLMs are generally capable of correctly identifying vulnerable functions, the quality and comprehensiveness of the explanations they provide remain more variable, especially in more complex code.

\begin{table*}[t]
\caption{Model Performance by Programming Language}
\label{tab:model-performance-language}
\centering
\begin{tabular*}{\textwidth}{@{\extracolsep{\fill}} l l c cccc cccc c}
\toprule
\textbf{Language} & \textbf{Model} & \textbf{N}
& \multicolumn{4}{c}{\textbf{Correctness}}
& \multicolumn{4}{c}{\textbf{Comprehensiveness}}
& \textbf{Overall Mean} \\
\cmidrule(lr){4-7} \cmidrule(lr){8-11}
& & 
& Mean & 0 & 1 & 2
& Mean & 0 & 1 & 2
& \\
\midrule
\multirow{4}{*}{C}
& Claude Haiku 4.5      & 173 & 1.942 & 3  & 4 & 166 & 1.902 & 2  & 13 & 158 & \cellcolor{bestgreen}\textbf{1.922} \\
& Gemini 3 Flash & 161 & 1.957 & 1  & 5 & 155 & 1.789 & 1  & 32 & 128 & 1.873 \\
& GPT-4.1 Mini   & 173 & 1.850 & 11 & 4 & 158 & 1.838 & 10 & 8  & 155 & 1.844 \\
& GPT-5 Mini     & 170 & 1.859 & 10 & 4 & 156 & 1.835 & 9  & 10 & 151 & 1.847 \\
\midrule
\multirow{4}{*}{Python}
& Claude Haiku 4.5      & 63 & 1.984 & 0 & 1 & 62 & 1.984 & 0 & 1 & 62 & \cellcolor{bestgreen}\textbf{1.984} \\
& Gemini 3 Flash & 61 & 1.934 & 1 & 2 & 58 & 1.852 & 1 & 7 & 53 & 1.893 \\
& GPT-4.1 Mini   & 63 & 1.794 & 6 & 1 & 56 & 1.778 & 6 & 2 & 55 & 1.786 \\
& GPT-5 Mini     & 63 & 1.841 & 5 & 0 & 58 & 1.841 & 5 & 0 & 58 & 1.841 \\
\midrule
\multirow{4}{*}{C++}
& Claude Haiku 4.5      & 15 & 2.000 & 0 & 0 & 15 & 2.000 & 0 & 0 & 15 & \cellcolor{bestgreen}\textbf{2.000} \\
& Gemini 3 Flash & 15 & 2.000 & 0 & 0 & 15 & 1.933 & 0 & 1 & 14 & 1.967 \\
& GPT-4.1 Mini   & 15 & 2.000 & 0 & 0 & 15 & 2.000 & 0 & 0 & 15 & \cellcolor{bestgreen}\textbf{2.000} \\
& GPT-5 Mini     & 14 & 2.000 & 0 & 0 & 14 & 2.000 & 0 & 0 & 14 & \cellcolor{bestgreen}\textbf{2.000} \\
\bottomrule
\end{tabular*}
\end{table*}

\vspace{2mm}
\begin{opus-box}
\textbf{Answering RQ4:} The four evaluated models demonstrated high explanatory quality, with overall mean scores above 1.83. Claude Haiku 4.5 leads in both evaluated dimensions, achieving a perfect score in 93.6\% of the assessments. The GPT models score 0 in correctness, with 6.1–6.8\% on the manually evaluated samples, compared to less than 1.2\% for Claude Haiku 4.5 and Gemini 3 Flash.
\end{opus-box}

\section{Discussion}

Three principal findings emerge from this study. First, the effect of interprocedural context depends on the model architecture. Second, a clear asymmetry exists between inference cost and performance gains: doubling the token count by adding additional context does not yield proportional improvements. 
Third, detection quality and explanation quality are coupled but not identical. We discuss below each finding in detail and examine its practical implications.

\subsection{Consequences of Including Interprocedural Context}

The impact of interprocedural context on detection accuracy varies across the LLMs. For Claude Haiku 4.5 and Gemini 3 Flash, neither form of additional context (CC or CK) produced performance variations that reached statistical significance across all languages. This indicates that these two models can extract the necessary information to detect vulnerabilities from a single function and remain robust to context variations. On the other hand, for GPT-4.1 Mini in C, the inclusion of CC context reduced accuracy from 75.6\% to 51.0\%, suggesting that including callee context introduces noise that degrades model performance. Similarly, incorporating caller context reduced accuracy to 68.8\%, reflecting the degradation pattern observed with CC.

For GPT-5 Mini in C, the effect was smaller but still statistically significant: with CK context, accuracy decreased from 90.52\% to 79.46\%. One hypothesis for this behavior is that GPT-family models, when supplied with additional context, expand their attention window to elements that are not directly relevant to the vulnerability. In contrast, models such as Claude Haiku 4.5 and Gemini 3 Flash appear more selective in incorporating additional context.

In Python, no comparison reached statistical significance for any model. One explanation is that the evaluated Python functions are more self-contained and exhibit lower interprocedural coupling than their C counterparts. For C++, the number of samples extracted from the dataset was insufficient to support reliable conclusions.

\subsection{Cost Variation}

The cost analysis revealed differences across LLMs. The total cost per model was: \$0.87 for GPT-4.1 Mini; \$3.04 for GPT-5 Mini; \$3.76 for Gemini 3 Flash; and \$7.23 for Claude Haiku 4.5, each for all context configurations. Thus, Claude Haiku is 8.3 times more expensive than GPT-4.1 Mini for executing the same tasks.

In terms of token consumption, the CO configuration in C required an average of 2,367 tokens per inference. Adding callees (CC) increased this to 4,785 tokens (+102\%), and adding callers (CK) to 4,586 tokens (+94\%). In other words, interprocedural context nearly doubles token usage and, consequently, cost. When cost is compared with performance, Gemini 3 Flash emerges as the best cost-performance balance: F1$\geq0.952$ across all tested configurations and languages at a total cost of \$3.76, lower than Claude Haiku 4.5 (\$7.23), which, although leading in explanation quality (RQ4), demonstrates equivalent performance in detection. GPT-4.1 Mini offers the lowest absolute cost (\$0.87), but its detection accuracy in C is lower than that of the other evaluated models.

\subsection{Explanation Quality}

The manual evaluation of explanations (RQ4) revealed that the LLMs can produce high-quality justifications for their vulnerability-detection decisions. Claude Haiku 4.5 led overall, with a mean score of 1.942/2.0 and 93.6\% of evaluations receiving the maximum score (2,2) across the assessed criteria, with only 1.2\% zero-scores in correctness. Gemini 3 Flash achieved a mean of 1.884/2.0 and 81.9\% perfect scores, but exhibited a notable gap between correctness (1.954) and comprehensiveness (1.814), a difference of 0.14 points. This suggests that while the model correctly identifies vulnerabilities, it sometimes fails to provide a coherent explanation of attack vectors and repair strategies.

GPT-5 Mini and GPT-4.1 Mini showed similar performance (means of 1.854 and 1.839, respectively), but with higher zero-score rates (6.1\% and 6.8\% in correctness), indicating that they commit fundamental identification errors more frequently than the other LLMs. The error pattern is consistent with the behavior observed in RQ1–RQ3: models that perform worse in detection (GPT variants) also exhibit more frequent explanatory errors. This suggests that explanation quality largely reflects underlying detection quality: LLMs that construct an internally coherent representation of the vulnerability tend to explain it more effectively.

\subsection{Practical Implications}

When analyzing all evaluation dimensions jointly, three implications emerge. First, the effect of interprocedural context is neither universally beneficial nor uniformly detrimental; it depends fundamentally on the model architecture. Prompt engineering strategies developed for one model should not be directly extrapolated to others, as their sensitivity to contextual expansion varies significantly.

Second, there is a clear asymmetry between inference cost and performance gains. The near doubling of token consumption caused by additional context does not translate into proportional improvements in accuracy and, in the case of GPT-4.1 Mini, results in statistically significant performance regression. This finding challenges the common assumption that ``more context is always better" in the development of LLM-based vulnerability detection systems.

Third, the evaluated models demonstrate that detection and explanation are coupled but non-identical capabilities. Claude Haiku 4.5 leads in explanatory quality, Gemini 3 Flash leads in cost-effective detection performance, and no single model dominates across all dimensions simultaneously. Consequently, selecting the optimal model depends on application priorities: explanatory completeness (Claude Haiku 4.5), cost–accuracy balance (Gemini 3 Flash), or minimal cost with reduced accuracy (GPT-4.1 Mini).

\section{Threats to Validity}

In this section, we discuss the threats to the validity of our study, and how we mitigate them~\cite{Wohlin2000}.

\textbf{Construct Validity:}
\textit{Training data contamination:} The evaluated LLMs were trained on large corpora collected from the web, including CVE reports and security patches. If a model was exposed to the ReposVul entries during training, its ability to correctly classify vulnerable functions may reflect memorization rather than genuine reasoning. We cannot guarantee the complete absence of overlap with the training data, as the data used to train these models is not publicly disclosed by their providers.
\textit{Manual evaluation:} The manual evaluation conducted for RQ4 relied on a rubric-based scoring scheme (0--2 scale for correctness and comprehensiveness) applied by a single practitioner across all LLM-generated explanations. Although the rubric was defined prior to scoring and piloted on a small sample to stabilize interpretation, the assessments are inherently subject to individual judgment.

\textbf{Internal Validity:}
\textit{Prompt engineering:} All prompts follow a fixed structure composed of a task instruction, a code block, and an interprocedural context block. We did not perform model-specific prompt optimization (prompt tuning). Different models may benefit from alternative formulations, and the reported results may therefore underestimate the potential performance of some models under optimized prompts.

\textbf{External Validity:}
\textit{Single dataset:} The entire study is based on ReposVul. Although this dataset is diverse in terms of projects (1,491 repositories) and vulnerability types (236 CWE types), the results may not generalize to other data sources, such as synthetic datasets, proprietary codebases, or vulnerabilities in other programming languages.
\textit{Model selection:} The four selected models represent lower-cost variants from three major providers, prioritizing long-context support and cost competitiveness. Higher-capacity models may exhibit different patterns of context sensitivity.
\textit{Used subset:} Dataset filtering retained only entries containing at least one vulnerable function (\texttt{target~=~1}). The absence of negative examples (non-vulnerable functions) prevents the evaluation of false positive rates. It limits the interpretation of accuracy, which in this study is equivalent to recall. The results should not be extrapolated to indicate overall detection performance in balanced datasets.

\textbf{Conclusion Validity:}
\textit{Variable sample size per configuration:} The number of evaluated functions in each context configuration varies across models, as the extraction of callers and callees depends on the information available in the call graph. We controlled for this variation by applying McNemar's test on paired subsets, isolating the effect of context manipulation and reducing biases arising from unequal sample sizes.
\textit{Reduced sample size in C++:} After filtering for interprocedural availability, the C++ subset contains 5--18 functions per configuration, with N = 1 in some cases. These sample sizes preclude reliable statistical testing. C++ results were excluded from statistical analysis and should be interpreted descriptively only.

\section{Related Work}

The use of LLMs for automated vulnerability detection has received increasing attention in the literature. Zhou et al.~\cite{zhou25, zhou2024emerging} presented a literature review of LLM-based vulnerability detection and repair, concluding that most existing studies have assessed LLMs on single-function benchmarks and that cost and generalizability remain underexplored. Germano et al.~\cite{germano25} find that explanation quality is rarely evaluated and that studies with multiple programming languages are uncommon. Our study directly addresses these gaps by evaluating cost alongside accuracy across multiple languages and including a manual assessment of explanation quality.

Ding et al.~\cite{ding2024primevul} evaluated code-specific models, such as CodeBERT and CodeT5, for vulnerability detection. Their study focuses on single-function detection in C/C++ and does not address interprocedural context or inference cost. Saimbhi et al.~\cite{saimbi24} applied the GPT-3.5 Turbo to detect vulnerabilities in PHP code. Fu et al.~\cite{fu2023chatgpt} evaluated ChatGPT for vulnerability detection, classification, and repair, reporting limitations when applied to real-world code. Khare et al.~\cite{khare2025understanding} evaluated 16 LLMs on 5,000 code samples spanning Java and C/C++, observing a mean accuracy of 62.8\% and a mean F1 score of 0.71, with performance on real-world datasets 10.5 percentage points lower than on synthetic datasets. Li et al.~\cite{li24} investigated the limitations of function-level vulnerability detection when confronted with interprocedural vulnerabilities, introducing the InterPVD dataset and the VulTrigger tool. Their results indicate that 24.3\% of vulnerabilities in real-world repositories are interprocedural and that function-level detectors perform significantly worse on this subset, demonstrating that function granularity is insufficient to capture cross-procedural dependencies.

Our study differs from prior work in four aspects: (i) we treat interprocedural context level as an explicit experimental variable rather than assuming broader context is beneficial; (ii) we consider three languages (C, C++, and Python) across 509 real-world CVEs; (iii) our analysis quantify the cost vs. performance trade-off per model and context configuration based on actual token consumption; and (iv) we manually assess the correctness and quality of model-generated explanations across 986 samples.

\section{Conclusion}

This study empirically investigated the impact of interprocedural context on vulnerability detection using LLMs. Four modern commercial LLMs, Claude Haiku 4.5, GPT-4.1 Mini, GPT-5 Mini, and Gemini 3 Flash, were evaluated on 509 CVEs extracted from the ReposVul dataset, covering C, C++, and Python. Surprisingly, across the evaluated context-variation configurations, the configuration with less context achieved higher accuracy. For the GPT models in C, including callees or callers reduced accuracy by up to 25\%, suggesting that additional interprocedural context introduces noise that degrades the model's reasoning. In contrast, Claude Haiku 4.5 and Gemini 3 Flash demonstrated robustness to context variation, maintaining stable performance across configurations. The effect of interprocedural context also varied by programming language. In C, degradation was more pronounced for the GPT models, whereas in Python, a similar but less pronounced trend was observed.

From a cost perspective, incorporating interprocedural context approximately doubled token consumption and, consequently, inference cost, without delivering proportional gains in accuracy. Gemini 3 Flash provided the best cost–performance trade-off, achieving an F1 score close to 1.0 in C at a total cost of \$3.76. Claude Haiku 4.5 led in explanation quality, receiving the maximum score (2,2) in 93.6\% of manual evaluations. No single model dominated across all dimensions, indicating that application priorities should guide model selection: explanation completeness (Claude Haiku 4.5), cost–accuracy balance (Gemini 3 Flash), or minimal cost with reduced accuracy (GPT-4.1 Mini).

These findings have direct implications for the design of LLM-based vulnerability detection tools. Developers should calibrate the level of interprocedural context according to both the target language and the specific model in use. Furthermore, explanation quality correlates with detection quality: models that misclassify more frequently also produce less precise explanations. This suggests that investing in models with stronger reasoning capabilities benefits both classification and explanation performance.

\section*{Data availability}

All prompts, collected data, and complementary results are in the supplementary material~\cite{supplementary}.

\bibliographystyle{ACM-Reference-Format}
\bibliography{sample-base}


\end{document}